\documentclass[reprint,nofootinbib,aps,superscriptaddress,amsmath,amssymb,onecolumn,11pt]{revtex4-2}

\usepackage{graphicx}
\usepackage{dcolumn}
\usepackage{bm}
\usepackage{braket}
\usepackage{breqn}
\usepackage{mathrsfs}
\usepackage{color}

\usepackage[colorlinks=true,linkcolor=red,citecolor=blue]{hyperref}

\begin{document}

\title{Energy extraction via magnetic reconnection in magnetized black holes}

\author{Shao-Jun Zhang}
\email{sjzhang@zjut.edu.cn}
\affiliation{Institute for Theoretical Physics and Cosmology$,$ Zhejiang University of Technology$,$ Hangzhou 310032$,$ China}
\affiliation{United Center for Gravitational Wave Physics$,$ Zhejiang University of Technology$,$ Hangzhou 310032$,$ China}
\date{\today}

\begin{abstract}
The Comisso-Asenjo mechanism is a novel mechanism proposed recently to extract energy from black holes through magnetic reconnection of the surrounding charged plasma, in which the magnetic field plays a crucial role. In this work, we revisit this process by taking into account the backreaction of the magnetic field on the black hole's geometry. We employ the Kerr-Melvin metric to describe the local near-horizon geometry of the magnetized black hole. By analyzing the circular orbits in the equatorial plane, energy extraction conditions, power and efficiency of the energy extraction, we found that while a stronger magnetic field can enhance plasma magnetization and aid energy extraction, its backreaction on the spacetime may hinder the process, with a larger magnetic field posing a greater obstacle. Balancing these effects, an optimal moderate magnetic field strength is found to be most conducive to energy extraction. Moreover, there is a maximum limit to the magnetic field strength associated with the black hole's spin, beyond which circular orbits in the equatorial plane are prohibited, thereby impeding energy extraction in the current scenario.
\end{abstract}


\maketitle
\section{Introduction}

Black holes (BHs), known as the most compact and unique objects envisioned by Einstein's gravitational theory, are believed to play a crucial role in various high-energy astrophysical phenomena, such as gamma-ray bursts (GRBs) and relativistic jets observed in active galactic nuclei (AGNs). The energy emitted in these occurrences can stem from either the gravitational potential energy discharged as material falls into the BH or from the intrinsic energy of the BH itself. According to general relativity (GR) and BH thermodynamics, rotating BHs harbor significant amounts of extractable energy, with the potential to reach up to $0.29 M c^2$ \cite{Christodoulou:1970wf}, where $M$ represents the BH's mass and $c$ denotes the speed of light in vacuum. This substantial energy arises from the rotational energy of the BH, leading to an intriguing exploration of the mechanisms involved in harnessing such a considerable fraction of the BH's energy.

The first energy extraction mechanism was proposed by Penrose in 1969, and is now known as Penrose process \cite{Penrose:1969pc}. This process entails the division of a particle within the ergoregion into two distinct particles. Despite its theoretical appeal, the Penrose process is deemed impractical due to the necessity for the newly formed particles to exhibit relative velocities exceeding half the speed of light, a scenario rarely observed in actual astrophysical phenomena \cite{Bardeen:1972fi, Wald:1974kya}. Penrose’s seminal work has spurred physicists to investigate various alternative mechanisms for extracting energy from BHs. These alternatives include collisional Penrose process \cite{piran1975high}, superradiant scattering \cite{Teukolsky:1974yv}, Blandford-Znajek (BZ) process \cite{Blandford:1977ds}, and magnetohydrodynamic (MHD) Penrose process \cite{1990ApJ...363..206T}. Of these, the BZ process is currently recognized as the most promising method for interpreting GRBs \cite{Lee:1999se, Tchekhovskoy:2008gq, Komissarov:2009dn} and relativistic jets in AGNs \cite{McKinney:2004ka, Hawley:2005xs, Komissarov:2007rc, Tchekhovskoy:2011zx}.

Recently, Comisso and Asenjo proposed a novel mechanism for extracting energy from BHs by utilizing magnetic reconnection processes within the ergoregion of a rotating BH \cite{Comisso:2020ykg} (see also a prior exploratory study \cite{Koide:2008xr}). This mechanism involves the generation of anti-parallel magnetic field configurations near the equatorial plane due to the BH's rotation \cite{Parfrey:2018dnc, Komissarov:2005wj, East:2018ayf, Ripperda:2020bpz}. The magnetic field direction changes, forming current sheets that give rise to plasmoids through a disruptive plasmoid instability process \cite{comisso2016General, Uzdensky:2010ts, Comisso:2017arh}. These plasmoids facilitate rapid magnetic reconnection, converting magnetic energy into plasma kinetic energy before being expelled from the reconnection layer \cite{daughton2009Transition, bhattacharjee2009Fast}. The magnetic field lines are elongated by the BH's rotation, initiating the formation of new current sheets and repeating the reconnection process. During each reconnection event, the plasma in the current sheet splits into corotating and counterrotating components. The corotating part is accelerated, while the counterrotating part is decelerated. Analogous to Penrose process, energy extraction from the BH is achieved by absorbing the decelerated portion carrying negative energy into the BH, allowing the accelerated portion to escape with additional energy obtained from the BH's rotational energy.

Observations have verified the presence of diverse magnetic field scales surrounding BHs, supported by accretion matter or companion stars. In particular, the supermassive BH Sagittarius A* is associated with the magnetar SGR J1745-2900, and a strong magnetic field is detected near the event horizon of the BH in M87* \cite{Mori:2013yda, Kennea:2013dfa, Eatough:2013nva, Olausen:2013bpa, EventHorizonTelescope:2021srq}. The magnetic reconnection process in BHs is expected to be a common occurrence. Research by Comisso and Asenjo suggests that this mechanism may outperform the BZ process in specific circumstances, making it a promising avenue for extracting energy from BHs. Expanding on this fundamental research, the Comisso-Asenjo mechanism has been applied to various other rotating BHs and scenarios \cite{Wei:2022jbi, Liu:2022qnr, Carleo:2022qlv, Khodadi:2022dff, Li:2023nmy, Li:2023htz, Khodadi:2023juk, Shaymatov:2023dtt, Wang:2022qmg, Zhang:2024rvk, Chen:2024ggq}. 

In all the mentioned work, the backreaction of the surrounding magnetic fields (and also the plasma) on the BH geometry has been ignored, as they are typically considered insignificant compared to the BH's gravitational energy. Nevertheless, recent astronomical observations have identified situations where the strength of magnetic fields in the universe necessitates a consideration of their influence on spacetime geometry \cite{Olausen:2013bpa, EventHorizonTelescope:2021srq}. Further elaboration on the magnetic field surrounding BHs will be provided in the main text. In any case, it is always of theoretical interest to consider the backreaction of the magnetic field on the BH geometry and, in turn, on the Comisso-Asenjo mechanism. 

Inspired by these works, we reexamine the Comisso-Asenjo mechanism by incorporating the influence of the magnetic field's backreaction on the geometry of the BH. Analyzing the dynamic interplay between BHs and surrounding magnetic fields is in general an intricate task, often requiring relativistic magnetohydrodynamic simulations. A more straightforward approach involves investigating stationary magnetized BH solutions to establish fundamental insights. Notably, solutions to the Einstein-Maxwell equations, such as Kerr-Melvin BHs within a uniform magnetic field along the symmetry axis, have been identified in the literature \cite{Wald:1974np, Ernst:1976mzr, Ernst:1976bsr, Gibbons:2013yq}. Recent studies have delved into the thermodynamic properties and astrophysical implications of Kerr-Melvin BHs, contributing significantly to our understanding of these BHs \cite{Budinova:2000yd, Bicak:2006hs, Gibbons:2013dna, Astorino:2016hls, Booth:2015nwa, Astorino:2015naa, Astorino:2016ybm, Gao:2022ckf, Wang:2021ara, Hou:2022eev, Chakraborty:2024aug}. We intend to utilize this simplified model to describe the local near-horizon geometry of the magnetized BHs, aiming to gain insights into the underlying physics. A thorough examination reveals that while increasing the magnetic field strength can improve plasma magnetization and aid in energy extraction, its influence on spacetime could impede the process, particularly with larger magnetic fields presenting a greater challenge. 

The paper is organized as follows. In Section II, we give a brief overview of the Kerr-Melvin BHs, along with an examination of how the magnetic field affects the ergoregion. Section III delves into a thorough analysis of the circular geodesic motion of the plasma in the equatorial plane. Section IV investigates the energy extraction from the BHs via the Comisso-Asenjo mechanism. The last section is the summary and conclusions.


\maketitle
\section{Kerr-Melvin black holes}

We consider the Kerr-Melvin metric, which is an exact stationary and axisymmetric electrovacuum solution describing a rotating BH immersed in an external uniform magnetic field. The metric takes the form \cite{Ernst:1976mzr, Ernst:1976bsr, Aliev:1989wx, Aliev:1989wz, Gibbons:2013yq, Astorino:2015naa}
\begin{align}
    ds^2 =  \Sigma |\Lambda|^2 \left[-\frac{\Delta}{A} dt^2 + \frac{dr^2}{\Delta} + d \theta^2 \right] + \frac{A  \sin^2 \theta}{ \Sigma |\Lambda|^2}  \left(|\Lambda_0|^2 d\phi - \omega dt\right)^2,
\end{align}
where 
\begin{align}
    \Delta &= r^2 + a^2 - 2 M r, \quad \Sigma = r^2 + a^2 \cos^2\theta,\quad  A = (r^2 + a^2)^2 - \Delta a^2 \sin^2 \theta,\nonumber\\
   \Lambda &= 1 + \frac{B^2 \sin^2 \theta}{4} \frac{A}{\Sigma} - \frac{i}{2} a B^2 M \cos\theta \left(3 - \cos^2\theta + \frac{a^2 \sin^4 \theta}{\Sigma}\right),\nonumber\\
   \omega & = \frac{\alpha - \beta \Delta}{r^2 + a^2},\nonumber\\
   \alpha & = a (1- a^2 M^2 B^4),\nonumber\\
   \beta & = \frac{a \Sigma}{A} + \frac{a M B^4}{16} \Bigg(- 8 r \cos^2 \theta (3 - \cos^2 \theta) - 6 r \sin^4 \theta + \frac{2 a ^2 \sin^6 \theta}{A} \left[r(r^2+a^2) + 2 M a^2\right] \nonumber\\
   &\quad+ \frac{4 M a^2 \cos^2\theta}{A} \left[(r^2 + a^2)(3-\cos^2\theta)^2 - 4 a^2 \sin^2 \theta\right]\Bigg).
\end{align}
Here $M$ and $a$ are the BH's mass and spin parameters respectively, and $B$ is the asymptotic magnetic field strength. The additional factor $|\Lambda_0|^2 \equiv |\Lambda(\theta=0)|^2 = 1 + a^2 M^2 B^4$ is introduced in the metric to remove the canonical singularities on the polar axis \cite{Aliev:1989wx, Aliev:1989wz}. Note the metric is not asymptotically flat but resembles the Melvin magnetic universe \cite{Melvin:1963qx}. Nevertheless, we employ it to describe the local geometry near the BH and assume that the spacetime is still asymptotically flat. We do not have to worry about the explicit geometry far away from the BH as it is not relevant to the physics we are going to study. 

When $B=0$, the metric reduces to the Kerr metric exactly; when $a=0$, it reduces to the Schwarzshild-Melvin metric \cite{Ernst:1976mzr}. The detailed electromagnetic field configuration expression, which is lengthy and not pertinent to this study, is omitted here. For more details, refer to \cite{Ernst:1976mzr, Ernst:1976bsr, Aliev:1989wx, Aliev:1989wz, Gibbons:2013yq, Astorino:2015naa}. 

In this work, we utilize the units $c=G=4\pi \epsilon_0 =1$, where $c, G, \epsilon_0$ are the speed of light in vacuum, the Newton gravitational constant and the vacuum permittivity, respectively. Moreover, it is convenient to set the BH mass $M=1$ so that all quantities are measured in units of $M$. In this convention, one can define a characteristic magnetic field $B_M = 1/M$ associated to a spacetime curvature of the same order of the horizon curvature. At this order, the electromagnetic energy is comparable to the gravitational energy of the BH \cite{Galtsov:1978ag, Aliev:1989wz}. Restoring physical units, we have
\begin{equation}
	B_M \sim 2.36 \times 10^{19} \left(\frac{M_\odot}{M}\right) \textrm{Gauss},
\end{equation}
where $M_\odot$ is the solar mass. $B_M$ is inversely proportional to the mass of the BH, and takes an extremely high value in general: For stellar-mass BHs with $M \sim 10 M_\odot$, $B_M \sim 10^{18}\textrm{Gauss}$; For intermediate-mass BHs with $M \sim 10^2 M_\odot$, $B_M \sim 10^{17}\textrm{Gauss}$; While for supermassive BHs with $M \sim 10^6 M_\odot$ (for example the Sagittarius A$^\ast$), $B_M \sim 10^{13} \textrm{Gauss}$. In our universe, the most powerful magnetic field observed to date emanates from the surface of magnetars, reaching up to $B \sim 10^{16} {\rm Gauss}$ \cite{Olausen:2013bpa}. While magnetic fields around BHs are typically orders of magnitude lower than that of neutron stars, they are not necessarily negligible compared to $B_M$, especially for supermassive BHs. Moreover, it is always of theoretical interest to consider extreme magnetic fields and their influences on BH physics. However, as we will see later, for the process we are going to study to happen, the magnetic field should not be too strong.

\begin{figure}[!htbp]
	\includegraphics[width=0.8\textwidth]{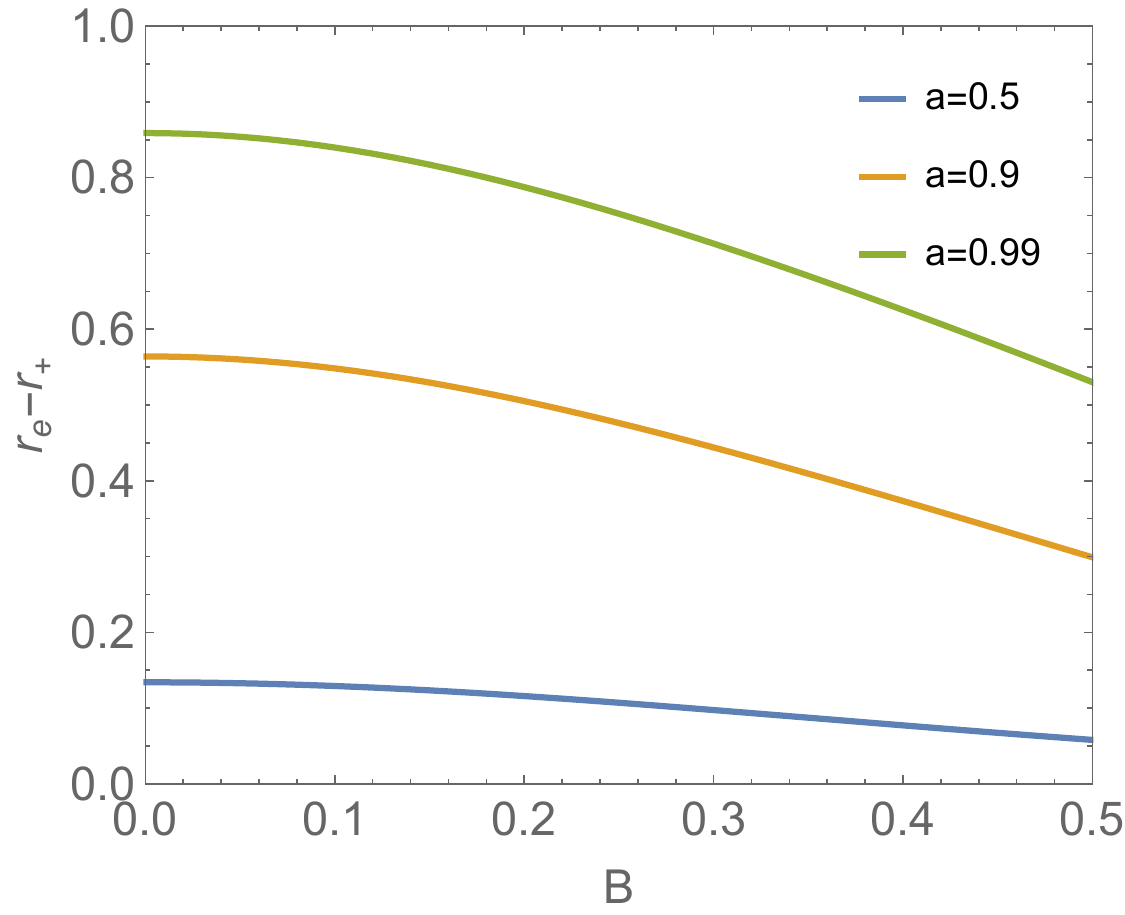}
	\caption{Ergosphere in the equatorial plane as a function of the magnetic field strength $B$ for various BH spin $a$. All physical quantities are measured in units of $M$.} \label{ErgoregionEquatorial}
\end{figure}

The location of the event horizon is not affected by the magnetic field and remains the same as that of the corresponding Kerr BH, that is, 
\begin{align}
    r_+ = M + \sqrt{M^2 - a^2}.
\end{align}
The radius of the ergosphere $r= r_e (\theta)$ is determined by the condition 
\begin{align}
    g_{tt} = - \Sigma |\Lambda|^2 \frac{\Delta }{A} + \frac{A \sin^2\theta}{\Sigma |\Lambda|^2} \omega^2 = 0.
\end{align}
The ergoregion is enclosed by the ergosphere and the event horizon, that is, $r_e > r >r_+$. In the limit of $B\rightarrow 0$, $r_e = M + \sqrt{M^2 - a^2 \cos^2 \theta}$. Our main focus in this work is on the ergoregion in the equatorial plane, i.e., $\theta = \frac{\pi}{2}$. In Fig. \ref{ErgoregionEquatorial}, we show the influence of the magnetic field on the shape of the ergoregion in the equatorial plane. From the figure, it can be seen that for a fixed BH spin $a$, larger $B$ will shrink the ergoregion. This implies that the backreaction of the magnetic field on spacetime is not conducive to the magnetic reconnection process.

\section{Circular orbits in equatorial plane}

Magnetic reconnection is associated with the movement of charged plasma in the vicinity of the compact object. Under the "force-free" assumption, the net electromagnetic force on the charged plasma is neglected, and thus the particles move on geodesics. With the two Killing vectors of the spacetime, $k \equiv \partial_t$ and $m \equiv \partial_\phi$, one can define two associated conserved quantities for the particles, the specific energy $E\equiv -g_{\mu\nu} k^\mu u^\nu$ and the $z$-component angular momentum $L_z \equiv g_{\mu\nu} m^\mu u^\nu$, with $u^\mu \equiv \dot{x}^\mu$ being the four-velocity of the particle (where dot means derivative with respect to some affine parameter). From the normalization condition of the four-velocity of the particle, i.e., $g_{\mu\nu} \dot{x}^\mu \dot{x}^\nu =-1$, we have the equation \cite{Zhang:2024rvk}
\begin{equation}
	g_{rr} \dot{r}^2 + g_{\theta \theta} \dot{\theta}^2 = V_{\rm eff} (r, \theta),
\end{equation}
where the effective potential $V_{\rm eff} (r, \theta)$ can be written in terms of $E$ and $L_z$ as
\begin{equation}\label{EffectivePotential}
	V_{\rm eff} (r, \theta)= \frac{g_{\phi \phi} E^2 + 2 g_{t \phi} E L_z + g_{tt} L_z^2}{g_{t \phi }^2-g_{tt} g_{\phi \phi }} - 1.
\end{equation}
For circular orbits in the equatorial plane, i.e., $\theta=\frac{\pi}{2}, \dot{\theta} =0$ and $\dot{r}=0$, the effective potential should satisfy the following conditions 
\begin{align}
	V_{\rm eff} =0,\quad \partial_r V_{\rm eff} =0,\quad \partial_\theta V_{\rm eff} =0.
\end{align}
The last condition can consistently be fulfilled due to the reflection symmetry of spacetime with respect to the equatorial plane. The first two conditions determine the Keplerian angular velocity $\Omega_K \equiv \dot{\phi} / \dot{t}$, the specific energy $E$, and the $z$-component angular momentum $L_z$ of the particle as a function of the radius of the circular orbit, that is \cite{Zhang:2024rvk}
\begin{align}
	\Omega_K &= \frac{-\partial_r g_{t \phi} \pm \sqrt{(\partial_r g_{t \phi} )^2 - (\partial_r g_{tt}) (\partial_r g_{\phi \phi}) } }{\partial_r g_{\phi \phi } },\label{AngularVelocity}\\
	E &= -\frac{g_{tt} + g_{t \phi} \Omega _K}{\sqrt{- g_{tt} - 2 g_{t \phi} \Omega _K - g_{\phi \phi} \Omega_K^2}},\label{SpecificEnergy}\\
	L_z &= \frac{g_{t \phi} + g_{\phi \phi} \Omega _K}{\sqrt{- g_{tt} - 2 g_{t \phi} \Omega _K - g_{\phi \phi} \Omega _K^2}},\label{AngularMomentum}
\end{align}
where the sign $+$ and $-$ stands for corotating and counterrotating orbits, respectively. 

\begin{figure}[!htbp]
	\includegraphics[width=0.8\textwidth]{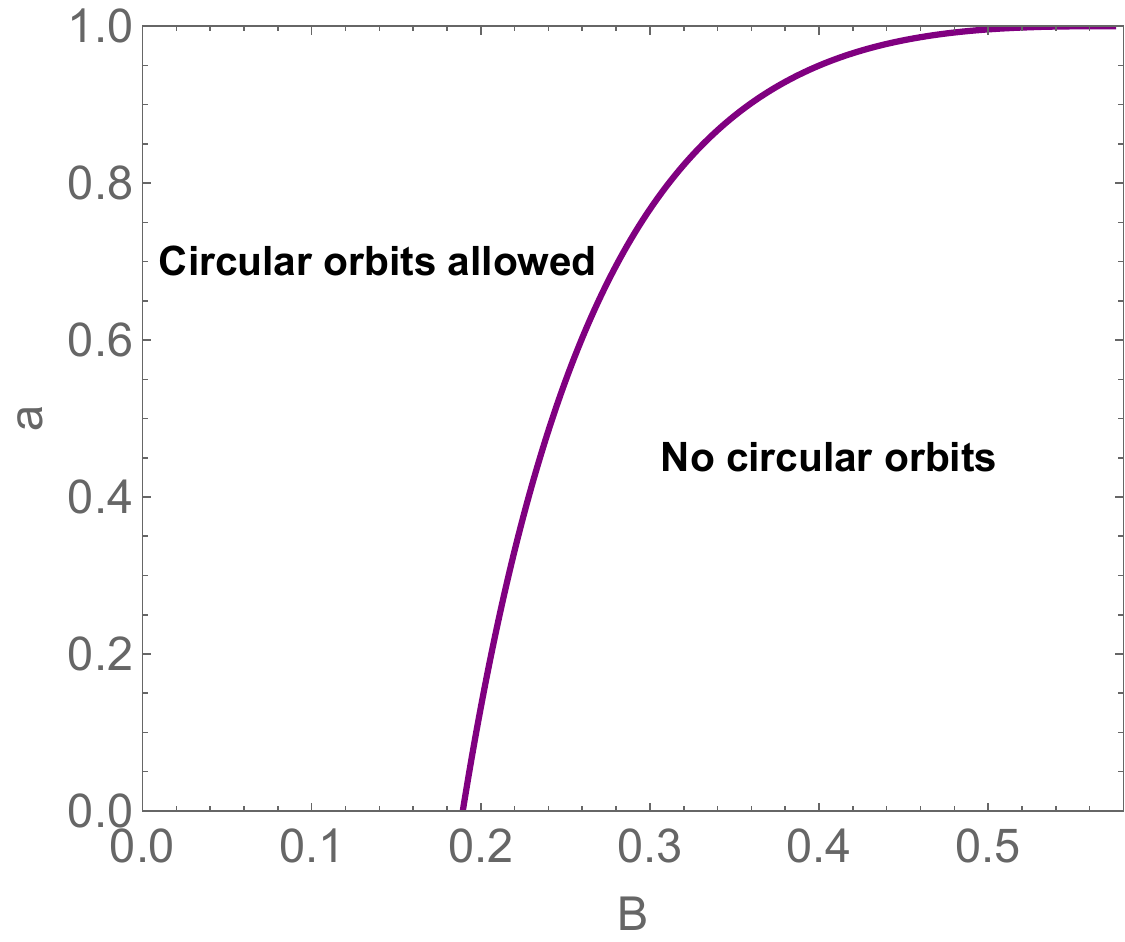}
	\caption{Allowed region in the $B-a$ plane for the existence of circular orbits in the equatorial plane. The two endpoints of the curve are $(a, B) \sim (0, 0.19)$ and $(1, 0.57)$, respectively. All physical quantities are measured in units of $M$.} \label{AllowedRegion}
\end{figure}

For circular orbits to exist, the above three orbital quantities (\ref{AngularVelocity}) (\ref{SpecificEnergy})(\ref{AngularMomentum}) should be real, so expressions under the root must be nonnegative. This physical requirement gives us an allowed region in the $B-a$ plane, as shown in Fig. \ref{AllowedRegion}. For a given $a$, from the figure, it can be seen that there exists an upper limit of the magnetic field strength $B=B_c$ over which no circular orbits exist, and $B_c$ increases along with the increase of $a$. In the non-rotating limit $a=0$, $B_c \sim 0.19$; While in the extremal limit $a \rightarrow 1$, $B_c$ does not diverge but approaches a finite value $B_c \sim 0.57$. This result implies that if the magnetic field is too strong, the circular orbits are prohibited, and thus the physical process we are considering will not occur. So in the following, we only consider $B \lesssim 0.57$.

\begin{figure}[!htbp]
	\includegraphics[width=0.8\textwidth]{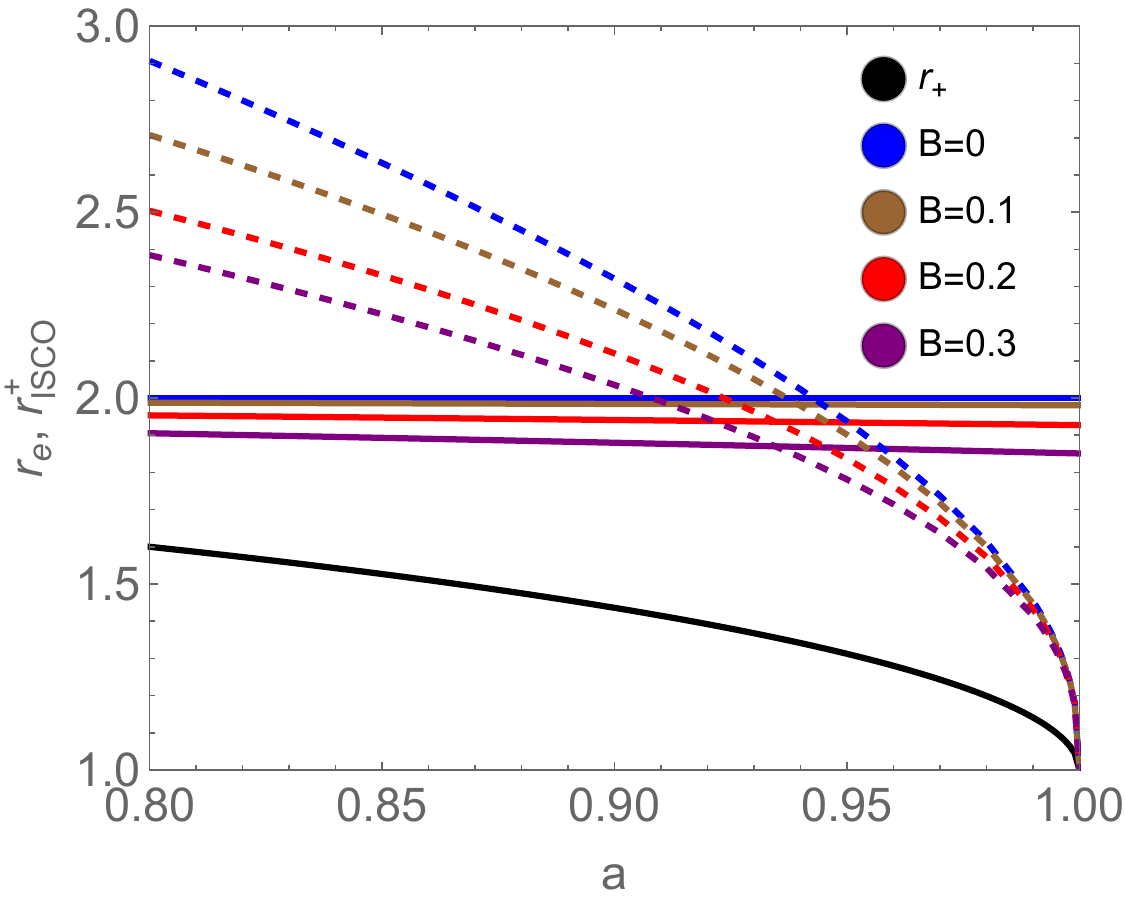}
	\caption{Corotating ISCOs $r_{\rm ISCO}^+$ as a function of $a$ for various $B$. Black solid cure is $r_+$, other solid ones are $r_e$, and dashed curves are $r_{\rm ISCO}^+$. All physical quantities are measured in units of $M$.} \label{ISCOsa}
\end{figure}

The counterrotating orbits are always outside the ergoregion, so we will only focus on the corotating ones. A radially stable circular orbit exists from infinity to the innermost (radially) stable circular orbit (ISCO), whose radius is determined by the condition
\begin{align}\label{ISCO}
	\partial_r^2 V_{\rm eff} = 0. 
\end{align}
In Fig. \ref{ISCOsa}, we plot the radius of the corotating ISCO $r_{\rm ISCO}^+$ as a function of $a$ for various $B$. From it, one can see that only when $a$ exceeds some extremely high value $a_c$ can $r_{\rm ISCO}^+$ enter the ergoregion. For example, $a_c \sim 0.943, 0.939, 0.934, 0.935$ for $B=0, 0.1, 0.2, 0.3$, respectively. It can be seen that with the increase of $B$, $a_c$ first decreases and then increases. In the extremal limit $a \rightarrow 1$, $r_{\rm ISCO}^+ \rightarrow r_+$ for any $B$. We have also numerically verified that the vertically stable condition $\partial_\theta^2 V_{\rm eff} \leq 0$ is consistent for the (radially) ISCO.

\section{Energy extraction via magnetic reconnection}

The Comisso-Asenjo mechanism facilitates energy extraction from BHs through magnetic reconnection occurring within the charged plasma situated in the ergoregion. Following \cite{Comisso:2020ykg}, we consider the plasma to rotate in a stable circular orbit in the equatorial plane at Keplerian velocity $\Omega_K$. Under the assumption of the one-fluid approximation, the energy-momentum tensor of the plasma takes the form
\begin{align}
	T^{\mu\nu} = p g^{\mu\nu} + \omega_0 U^\mu U^\nu + F^\mu_{~~\sigma} F^{\nu \sigma} - \frac{1}{4} g^{\mu\nu} F^{\rho\sigma} F_{\rho \sigma}, \label{EnergyMomentum}
\end{align} 
where $p, \omega_0, U^\mu$ and $F^{\mu\nu}$ are the proper plasma pressure, enthalpy density, four-velocity, and electromagnetic field tensors, respectively. With the time-like killing vector $\chi = \partial_t$, one can define a covariant conserved energy current $J^\mu \equiv T^{\mu \nu} \chi _\nu$ and the associated energy density,
\begin{align}
    e^{\infty} \equiv n_\mu J^\mu = - \alpha g_{\mu 0} T^{\mu 0},
\end{align}
where $n^\mu$ is the unit vector normal to time-like hypersurfaces $t=consant$. $e^{\infty}$ is usually called the ``energy-at-infinity'' density. 

To evaluate $e^{\infty}$, it is convenient to express it in terms of physical quantities in ``zero-angular-momentum-observer'' (ZAMO) frame $(\hat{t}, \hat{x}^1 = \hat{r}, \hat{x}^2 = \hat{\theta}, \hat{x}^3 = \hat{\phi})$ \cite{Bardeen:1972fi}. The ZAMO frame is a locally non-rotating frame in which the spacetime is locally Minkowskian, i.e., $ds^2 = - d \hat{t}^2 + \sum_{i=1}^3 (d \hat{x}^i)^2 = \eta_{\mu\nu} d\hat{x}^\mu d\hat{x}^\nu.$ It is related to the Boyer-Lindquist coordinates $(t, x^1 = r, x^2 = \theta, x^3 = \phi)$ by the transformations $d \hat{t} = \alpha dt$ and $d\hat{x}^i = \sqrt{g_{ii}} dx^i - \alpha \beta^i dt$, where the lapse function $\alpha$ and the shift vector $\beta^i = (0,0,\beta^\phi)$ are
\begin{align}
	\alpha = \left(- g_{tt} + \frac{g_{t\phi}^2}{g_{\phi\phi}}\right)^{1/2},\quad \beta^\phi = \frac{\sqrt{g_{\phi\phi}} \omega^\phi}{\alpha},
\end{align}
and $\omega^\phi \equiv - g_{t\phi} / g_{\phi\phi}$ is the angular velocity of the frame dragging. Quantities in ZAMO frame are denoted with hats. The Keplerian velocity of the co-rotating bulk plasma in ZAMO frame becomes 
\begin{align}
	\hat{v}_K = \frac{d\hat{x}^\phi}{d\hat{t}} = \frac{\sqrt{g_{\phi\phi}}}{\alpha} \Omega_K - \beta^\phi.
\end{align}

During each instance of magnetic reconnection, the plasma flowing out will divide into two segments, with one segment experiencing deceleration and the other acceleration. Assuming a significant conversion of magnetic energy to kinetic energy, allowing the electromagnetic energy density to be disregarded, and under the assumption of incompressible and adiabatic plasma, the energy density per enthalpy at infinity for the two segments can be expressed as follows \cite{Comisso:2020ykg}
\begin{align}
	\epsilon^\infty_\pm = \alpha \hat{\gamma}_K \left[(1+\beta^\phi \hat{v}_K) (1+\sigma_0)^{1/2} \pm \cos\xi(\beta^\phi + \hat{v}_K) \sigma_0^{1/2} - \frac{1}{4} 
\frac{(1+\sigma_0)^{1/2} \mp \cos\xi \hat{v}_K \sigma_0^{1/2}}{\hat{\gamma}^2_K (1+\sigma_0 - \cos^2\xi \hat{v}^2_K \sigma_0)}\right], \label{EnergyAtInfinity}
\end{align}
where the signs $+$ and $-$ stand for the accelerated and decelerated parts, respectively.  The angle $\xi$ represents the orientation between the velocity of the plasma outflow and the azimuthal direction in the equatorial plane, as observed in the local rest frame. Here, $\hat{\gamma}_K = (1 - \hat{v}_K^2)^{-1/2}$ and $\sigma_0$ denotes the plasma magnetization defined as
\begin{align}
    \sigma _0 \equiv \frac{B^2}{\omega_0}.
\end{align} 
From the above equation, it can be seen that $\epsilon^\infty_\pm$ is determined by a set of five parameters $\{a, B, \omega_0, \xi, r_X\}$, with $r=r_X$ being the radius of the circular orbit (X point).

For energy extraction from the BH to occur, specific energy extraction conditions must be met. These conditions necessitate that the decelerated portion exhibits negative energy-at-infinity, whereas the accelerated portion must manifest positive energy-at-infinity exceeding both its rest mass and thermal energy, that is
\begin{align}\label{EnergyExtractionConditions}
	\epsilon ^\infty_- <0, \quad \Delta \epsilon ^\infty_+ = \epsilon ^\infty_+ - \left(1 - \frac{\Gamma}{\Gamma -1} \frac{p}{\omega}\right) = \epsilon ^\infty_+ >0.
\end{align} 
Here plasma is assumed to be relativistically hot with polytropic index $\Gamma = 4 / 3$.

\subsection{Parameter space analysis}

Let us first analyze the influences of the magnetic field $B$ on the energy extraction conditions (\ref{EnergyExtractionConditions}).

\begin{figure}[!htbp]
	\includegraphics[width=0.8\textwidth]{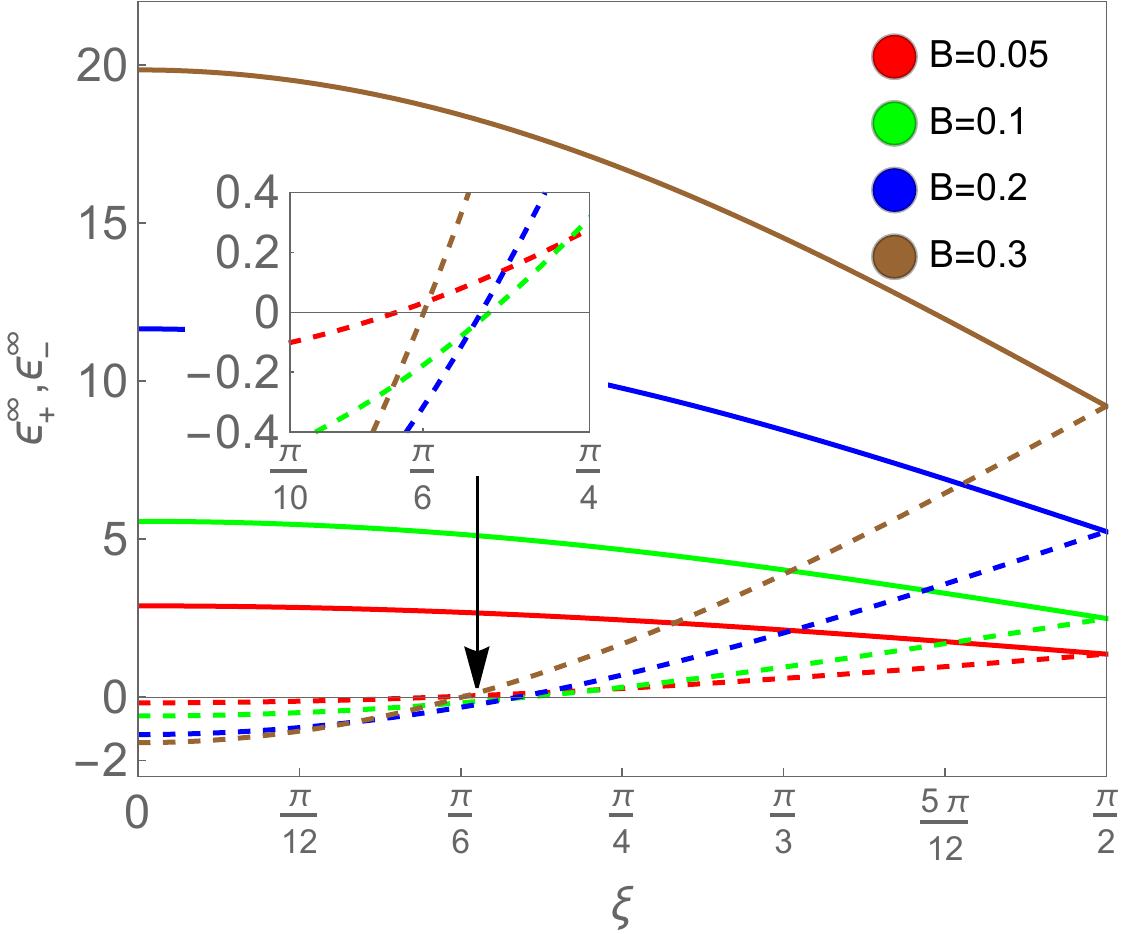}
	\caption{The energy-at-infinity density $\epsilon^\infty_\pm$ as a function of the orientation angle $\xi$ for various $B$, with $a=0.99, \omega_0=0.001$ and $r_X = r^+_{\rm ISCO}$. Solid curves are $\epsilon ^\infty_+$ while dashed ones are $\epsilon ^\infty_-$. All physical quantities are measured in units of $M$.} \label{EpsilonXi}
\end{figure}

\begin{figure}[!htbp]
	\includegraphics[width=0.8\textwidth]{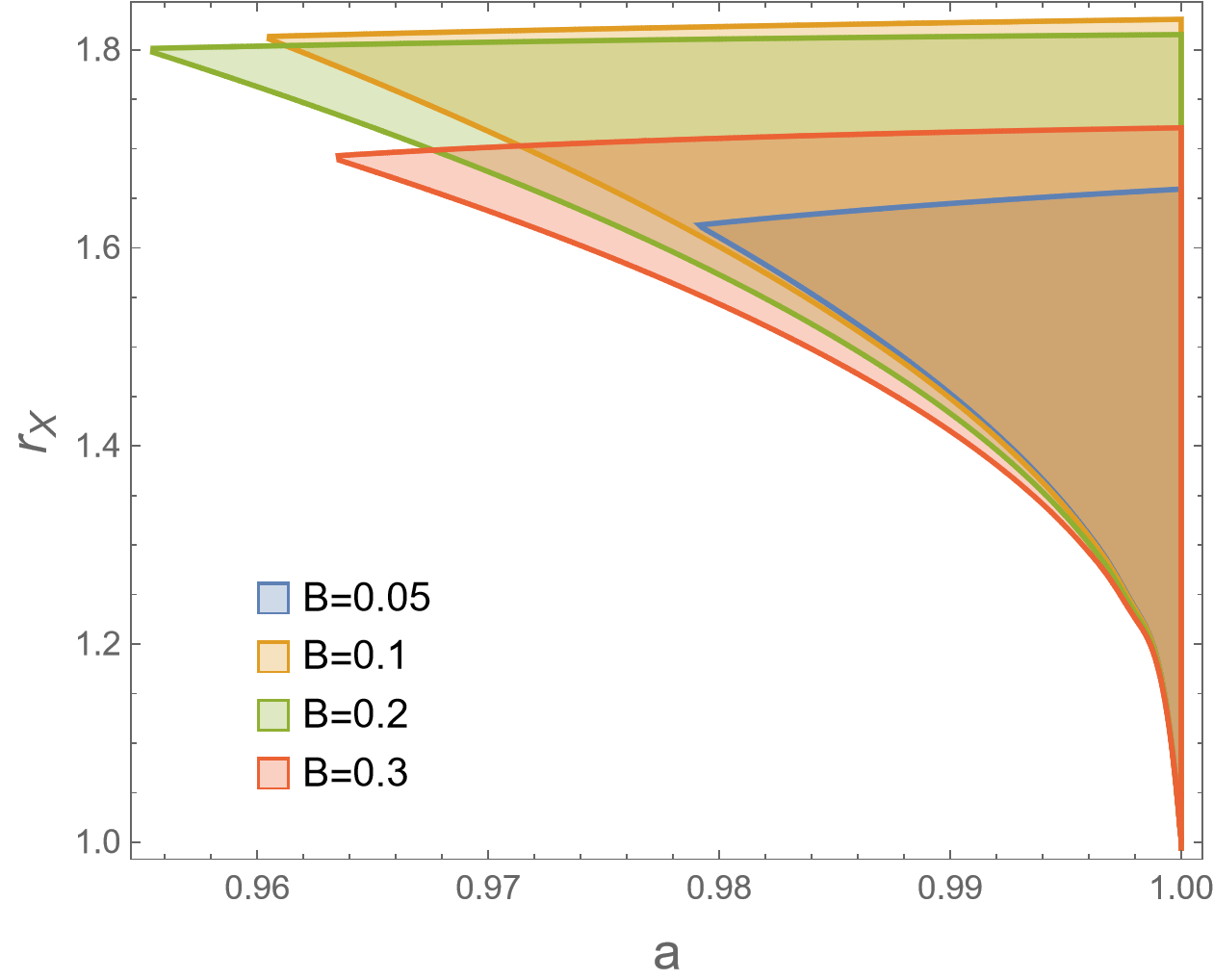}
	\caption{Allowed regions (shaded) in $a-r_X$ plane where the energy extraction conditions are satisfied for $r^+_{\rm ISCO}\leq r_X<r_{\rm e}$. We set $\xi=\frac{\pi}{12}$ and $\omega_0=0.001$. All physical quantities are measured in units of $M$.} \label{EpsilonraB}
\end{figure}

In Fig. \ref{EpsilonXi}, we show the influence of $B$ on the required orientation angle $\xi$ to satisfy the energy extraction conditions (\ref{EnergyExtractionConditions}) for a near-extreme BH ($a=0.99$). Without loss of generality, we set the enthalpy density $\omega_0=0.001$ and take the X-point to be at $r_X = r^+_{\rm ISCO}$. From the figure, one can see that for fixed $B$, $\epsilon ^\infty_+ > 0$ is always satisfied, while $\epsilon ^\infty_- < 0$ is only satisfied when $\xi$ is less than some upper bound $\xi ^c$. This is similar to the Kerr case \cite{Comisso:2020ykg}, which implies that energy extraction is favored by smaller $\xi$. As $B$ increases, $\xi^c$ first increases and then decreases. This observation indicates that optimal energy extraction occurs when $B$ takes moderate strength, while energy extraction becomes harder for too small or too large $B$.

In Fig. \ref{EpsilonraB},  we plot the allowed regions in the $a-r_X$ plane where the energy extraction conditions (\ref{EnergyExtractionConditions}) are met for various $B$. From the figure, one can see that as $B$ increases, the allowed region first expands and then shrinks, which once again indicates that a moderate $B$ is most favorable for energy extraction. 

\subsection{Energy extraction power and efficiency}

To evaluate the feasibility of energy extraction via magnetic reconnection, it is necessary to compute both the power and efficiency of the extraction process. The power ${\cal P}_{\rm extr}$ per unit enthalpy extracted from the BH can be well estimated as \cite{Comisso:2020ykg}
\begin{align}
	{\cal P}_{\rm extr} = - \epsilon ^\infty_- A_{\rm in} U_{\rm in},
\end{align}
while the efficiency is defined as \cite{Comisso:2020ykg}
\begin{align}
	\eta = \frac{\epsilon^\infty_+}{\epsilon^\infty_+ + \epsilon^\infty_-}.
\end{align}
The cross-sectional area of the incoming plasma, denoted as $A_{\rm in}$, can be approximated as $A_{\rm in} \sim (r_{\rm e}^2 - r_{\rm ISCO}^2)$ for highly rotating BHs. The parameter $U_{\rm in}$ is on the order of $10^{-2}$ and $10^{-1}$ for the collisional and collisionless regimes, respectively. If $\eta>1$, energy is extracted from the BH.

\begin{figure}[!htbp]
	\includegraphics[width=0.45\textwidth]{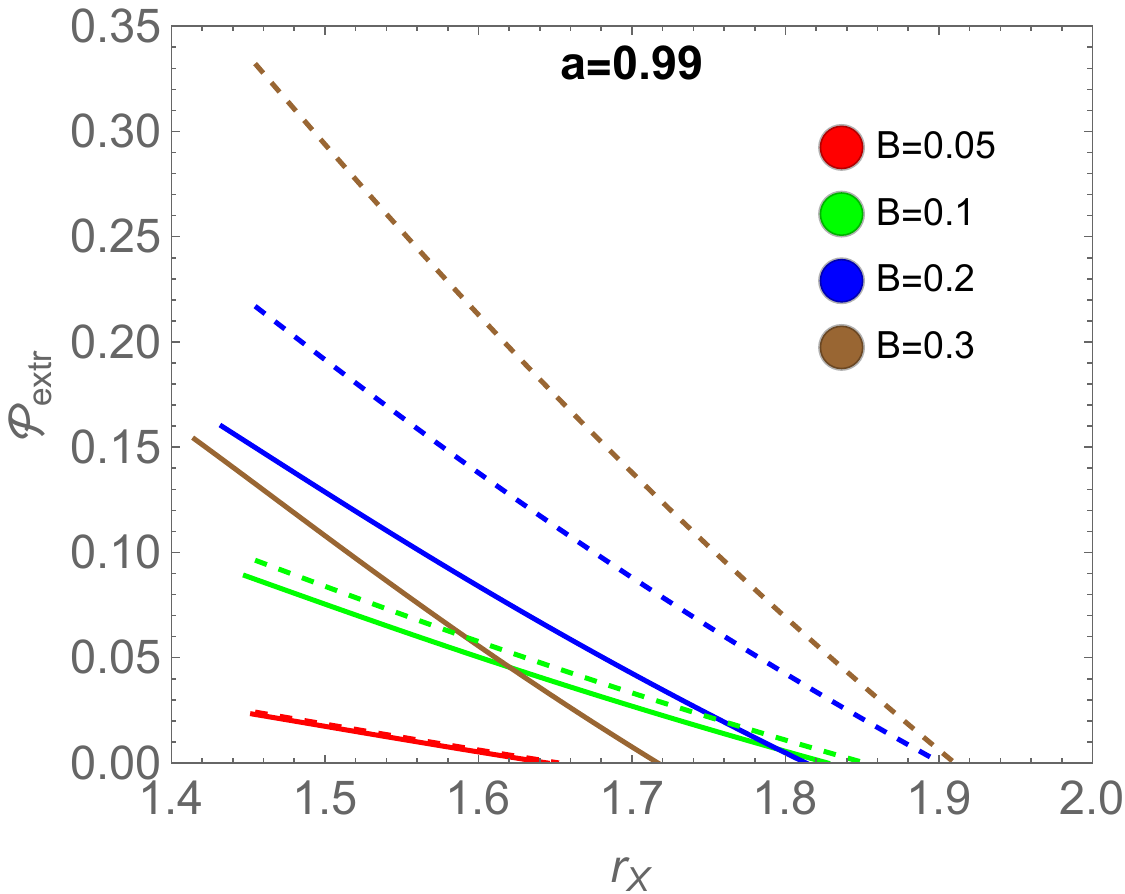}\quad
	\includegraphics[width=0.45\textwidth]{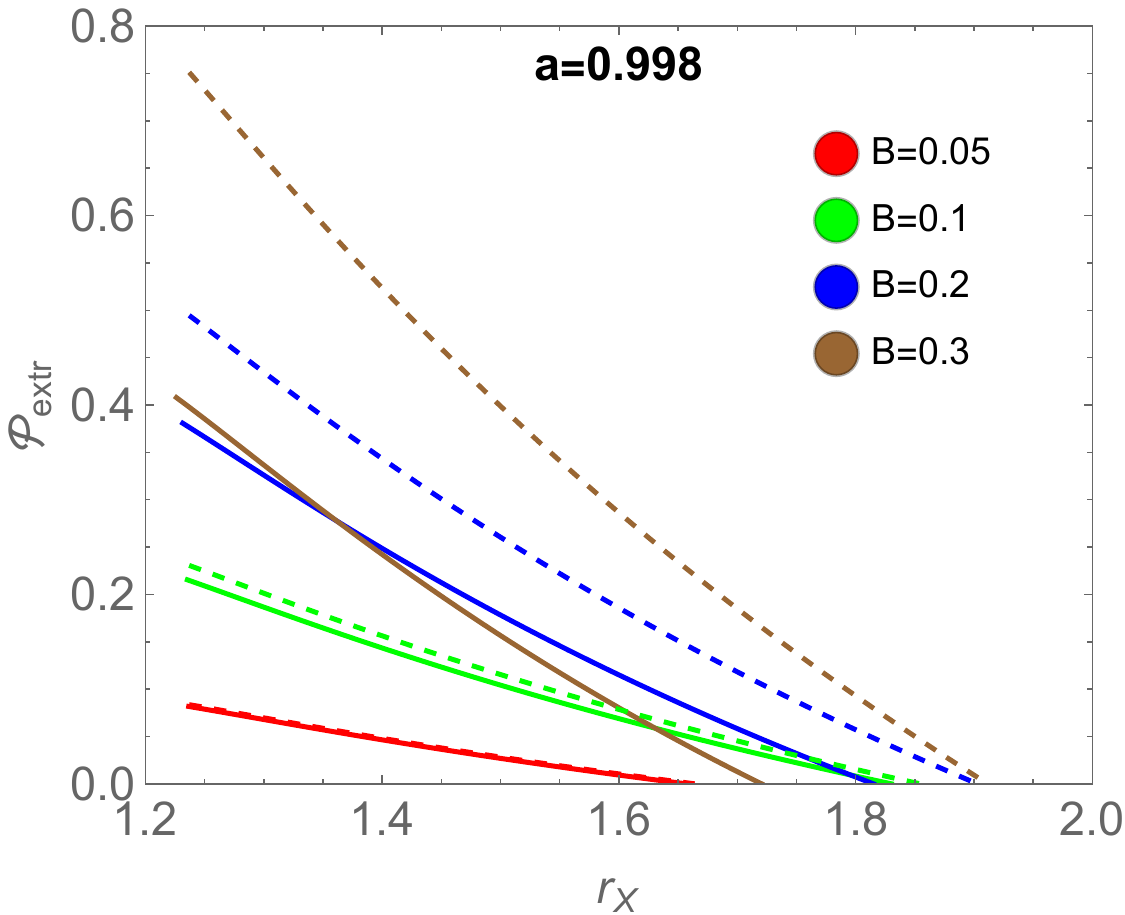}
	\caption{${\cal P}_{\rm extr}$ as a function of the X-point location $r_X$ for various $B$, with $\xi=\frac{\pi}{12}$ and $\omega_0=0.001$ and $U_{\rm in}=0.1$. $r_X$ is restricted to be in the range $r^+_{\rm ISCO}\leq r_X<r_{\rm e}$. Solid curves are for the Kerr-Melvin case, while dashed ones are the corresponding Kerr case without considering the backreaction of the magnetic field on the spacetime. All physical quantities are measured in units of $M$.} \label{PowerR}
\end{figure}

\begin{figure}[!htbp]
	\includegraphics[width=0.45\textwidth]{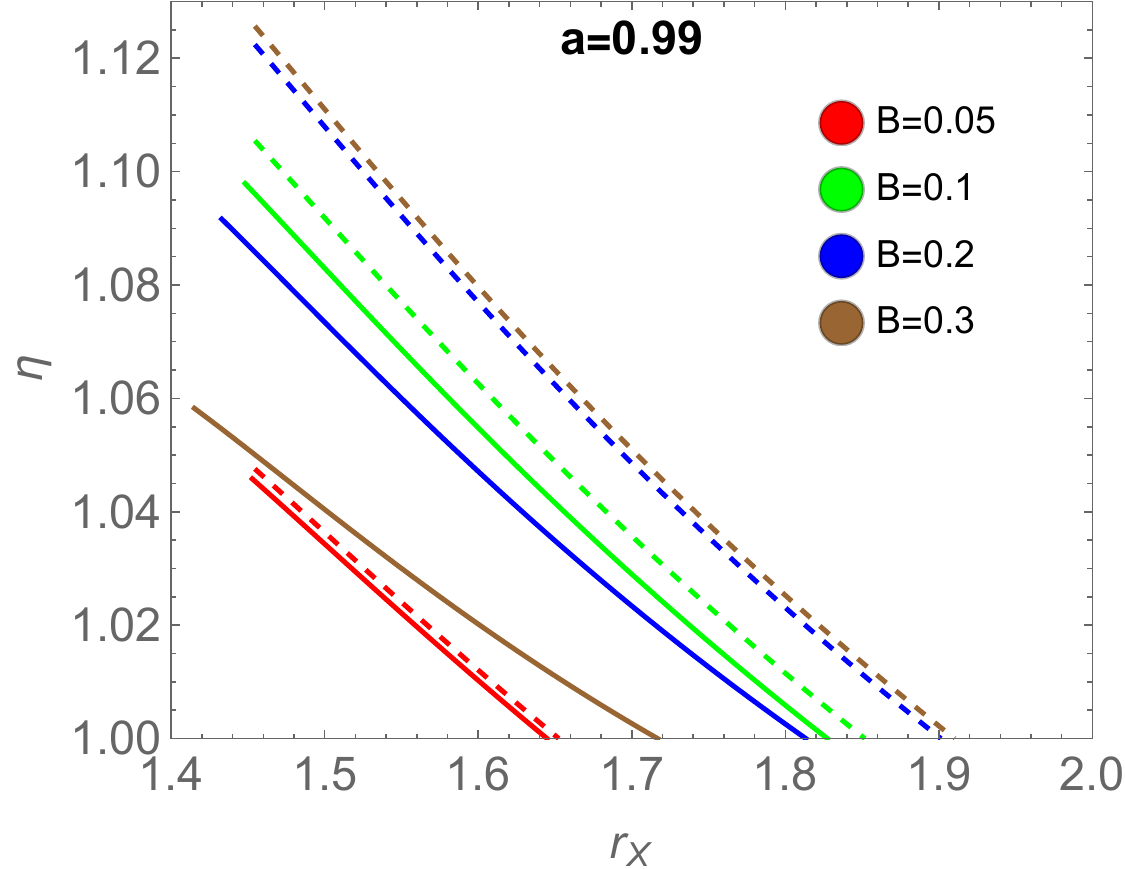}\quad
	\includegraphics[width=0.45\textwidth]{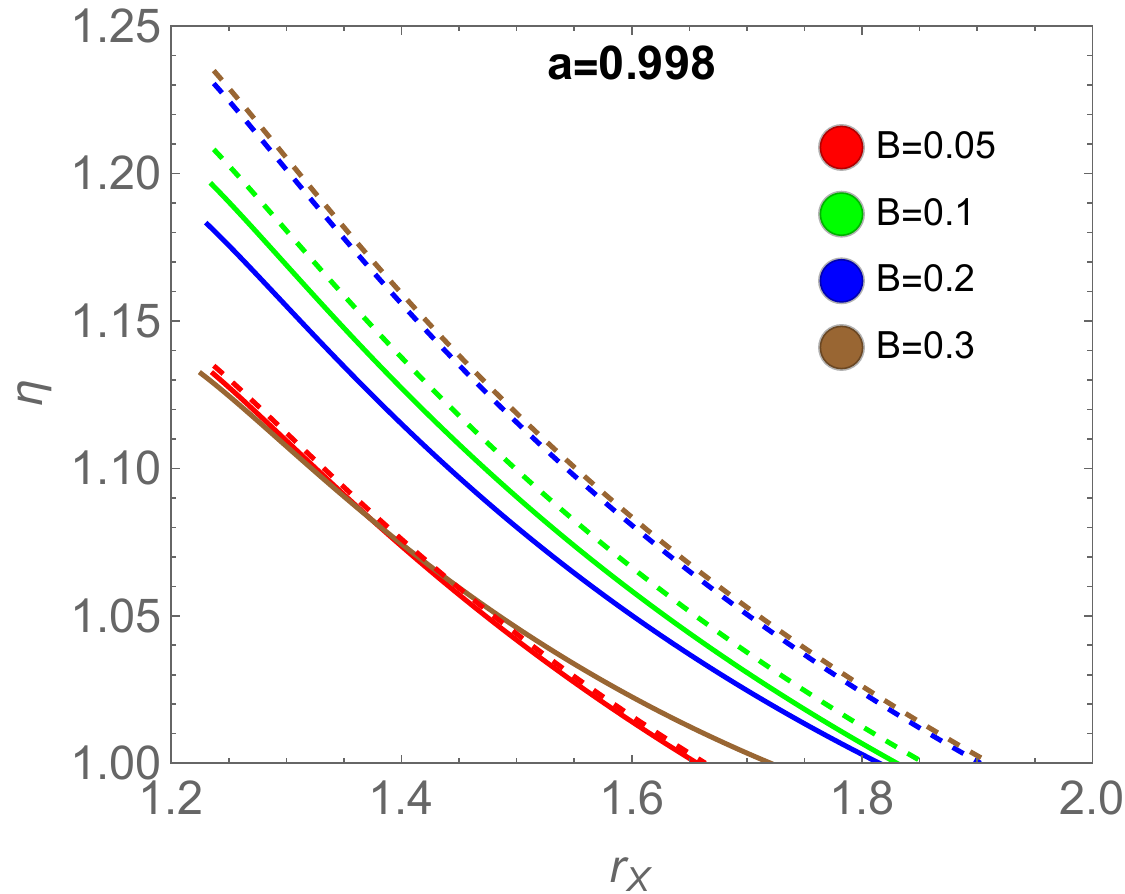}
	\caption{$\eta$ as a function of the X-point location $r_X$ for various $B$, with $\xi=\frac{\pi}{12}$ and $\omega_0=0.001$. $r_X$ is restricted to be in the range $r^+_{\rm ISCO}\leq r_X<r_{\rm e}$. Solid curves are for the Kerr-Melvin case, while dashed ones are the corresponding Kerr case without considering the backreaction of the magnetic field on the spacetime. All physical quantities are measured in units of $M$.} \label{EfficiencyR}
\end{figure}

In Fig. \ref{PowerR}, we show a typical picture of the power ${\cal P}_{\rm extr}$ as a function of $r_X$ for various $B$. For comparison, we also show the Kerr case where the backreaction of the magnetic field on the spacetime is not taken into account. From the figure it can be seen that, with all other parameters fixed, ${\cal P}_{\rm extr}$ is a monotonically decreasing function of $r_X$, reaching a maximum at $r_X=r^+_{\rm ISCO}$. As $B$ increases, ${\cal P}_{\rm extr}$ first increases and then decreases, which once again indicates that a moderate $B$ is most favorable for energy extraction. Compared with the Kerr case, we always have a lower power ${\cal P}_{\rm extr}$ in the Kerr-Melvin case. This signifies that the interaction of the magnetic field with spacetime has the potential to impede energy extraction. And the larger $B$, the stronger this impediment. For example, when $a=0.99$, the maximum power ratio in the two cases can reach $\frac{{\cal P}^{\rm max}_{\rm extr} ({\rm Kerr-Melvin})}{{\cal P}^{\rm max}_{\rm extr} ({\rm Kerr})} \sim 95\%, 93\%, 74\%, 47\%$ for $B=0.05, 0.1, 0.2, 0.3$, respectively. Similarly, for $a=0.998$, $\frac{{\cal P}^{\rm max}_{\rm extr} ({\rm Kerr-Melvin})}{{\cal P}^{\rm max}_{\rm extr} ({\rm Kerr})} \sim 96\%, 93\%, 77\%, 55\%$ for $B=0.05, 0.1, 0.2, 0.3$, respectively.

In Fig. \ref{EfficiencyR}, we show a typical picture of the efficiency $\eta$ as a function of the X-point location $r_X$ for various $B$. From the figure, it can be seen that the influence of $B$ on $\eta$ is similar to its influence on ${\cal P}_{\rm extr}$. As $B$ increases, $\eta$ first increases and then decreases, again indicating that a moderate $B$ is the most favorable for energy extraction. Compared with the Kerr case, we always have a smaller $\eta$ in the Kerr-Melvin case. This once again signifies that the interaction of the magnetic field with spacetime has the potential to impede energy extraction. And the larger $B$, the stronger this impediment. For example, when $a=0.99$, the maximum efficiency ratio in the two cases can reach $\frac{\eta^{\rm max} ({\rm Kerr-Melvin})}{\eta^{\rm max} ({\rm Kerr})} \sim 99.8\%, 99.3\%, 97.3\%, 94.0\%$ for $B=0.05, 0.1, 0.2, 0.3$, respectively. Similarly, for $a=0.998$, $\frac{\eta^{\rm max} ({\rm Kerr-Melvin})}{\eta^{\rm max} ({\rm Kerr})} \sim 99.7\%, 99.1\%, 96.2\%, 91.7\%$ for $B=0.05, 0.1, 0.2, 0.3$, respectively.

\section{Summary and conclusions}

In this study, we revisit the Comisso-Asenjo mechanism \cite{Comisso:2020ykg} by considering the backreaction of the magnetic field on spacetime. The analysis focuses on a fundamental model where the Kerr-Melvin metric describes the local geometry near the horizon. By studying circular orbits in the equatorial plane, evaluating energy extraction conditions, power and efficiency of the energy extraction, we found the significant influence of the backreaction on the process. 

The magnetic field $B$ exerts a significant influence on the geometry of BHs, impacting both the ergoregion and the surrounding orbits. A stronger magnetic field results in a reduction of the ergoregion in the equatorial plane where magnetic reconnection occurs, as illustrated in Fig. \ref{ErgoregionEquatorial}. There is a critical value for the magnetic field strength $B=B_c$, beyond which circular orbits cease to exist in the equatorial plane, as shown in Fig. \ref{AllowedRegion}. The value of $B_c$ rises with the BH spin $a$, and in the extreme limit $a \rightarrow 1$, $B_c \sim 0.57$. Consequently, if $B > 0.57$, magnetic reconnection cannot take place in the current scenario. As shown in Fig. \ref{ISCOsa}, only when $a$ exceeds some extremely high value $a_c$ can $r_{\rm ISCO}^+$ enter the ergoregion. For example, $a_c \sim 0.943, 0.939, 0.934, 0.935$ for $B=0, 0.1, 0.2, 0.3$, respectively. It can be seen that with the increase of $B$, $a_c$ first decreases and then increases.

This effect of magnetic fields on spacetime geometry further affects energy extraction. As $B$ increases, the required minimal orientation angle $\xi$ to satisfy the energy extraction conditions (\ref{EnergyExtractionConditions}) first increases and then decreases, as shown in Fig. \ref{EpsilonXi}. Additionally, with increasing $B$, the allowed region in the $a-r_X$ plane to meet the energy extraction conditions first expands and then shrinks, as shown in Fig. \ref{EpsilonraB}.

Upon analyzing the power ${\cal P}_{\rm extr}$ and efficiency $\eta$ of energy extraction as illustrated in Figs. \ref{PowerR} and \ref{EfficiencyR}, it was observed that both quantities exhibit an initial increase followed by a decrease with the rise of $B$. The values consistently remain noticeably lower than their Kerr counterparts, showing a more significant decrease as $B$ increases. 

All these results imply that while a stronger magnetic field enhances plasma magnetization, thereby promoting energy extraction, its backreaction on spacetime poses challenges. The interplay between these factors indicates that a moderate magnetic field strength is most conducive to energy extraction. Notably, there is a maximum limit for the magnetic field strength $B_c$ linked to the BH spin parameter $a$, beyond which circular orbits are restricted, hindering energy extraction in the current scenario. Nevertheless, it is important to acknowledge that energy extraction may still be achieved through alternative orbits like elliptic orbits or those not confined to the equatorial plane.

The backreaction of the magnetic field on spacetime becomes noticeable when it reaches a strength comparable to $B_M$, which is typically very high. Notably, quantum effects become important at the critical Schwinger field of $B_{\rm QED} \simeq 4.4 \times 10^{13} {\rm Gauss}$ (for references, see, e.g., \cite{Duncan:2000pj,vanPutten:1999tp}), necessitating their consideration in our analysis when the magnetic field $B$ approaches the magnitude of $B_{\rm QED}$. \footnote{We thank the referee for pointing this out.}

We utilized the Kerr-Melvin metric in this study to elucidate the backreaction of the magnetic field on the BH's geometry. While this model offers a simplified perspective, it aids in comprehending the influences of the magnetic field on the Comisso-Asenjo mechanism to a certain degree. However, in more realistic astrophysical scenarios, the incorporation of intricate models or relativistic magnetohydrodynamics is essential to depict the interplay between BHs and magnetic fields accurately, which calls for further investigations.

\begin{acknowledgments}

	This work is supported by the National Natural Science Foundation of China (NNSFC) under Grant No 12075207.

\end{acknowledgments}

\bibliographystyle{utphys}
\bibliography{refs}
\end{document}